# Arctic Dispersion Interruption Phenomenon and Sound Source Depth Estimation

Weng Jinbao[1]*, Yang Yanming[1], Chen Benqing[1], Xu Dewei[1], Zhou Hongtao[1]

1 Laboratory of ocean acoustic and remote sensing, Third Institute of Oceanography, Ministry of Natural Resources, Xiamen, China

## Abstract

The sound speed profile in the deep Arctic Ocean causes the surface layer to form a normal mode waveguide. When the depth of the sound source or receiver is near a node of the eigenfunction, the source cannot excite the mode, or the receiver cannot detect it, resulting in the modal amplitude at the receiver being approximately zero. This manifests as a dispersion interruption in the dispersion structure, which can be observed through time-frequency analysis of the received acoustic signal. Based on the interruption frequency identified from the received signal, combined with the relationship between node depth and modal frequency calculated from the ocean sound speed profile, the source depth can be estimated if the receiver depth is known. The phenomenon of dispersion interruption and the method of estimating source depth have been validated through simulations and experiments.

*wengjinbao@tio.org.cn

# I. INTRODUCTION

The underwater sound field in the Arctic exhibits unique characteristics due to the region's distinctive marine environment. Complex factors include intricate sound propagation environments and unique sound speed profiles. The complex propagation environment arises from sea-ice cover on the surface, while the unique sound speed profile is characterized by a monotonically increasing structure, and in some areas, a distinctive double-duct profile. These environmental factors have different impact mechanisms on sound propagation. Collectively, they result in underwater acoustic field effects in the deep Arctic that are completely different from those in conventional deep-sea and shallow-sea areas.

Underwater sound source localization has always been a challenging problem in underwater acoustics, especially for source depth estimation. Currently, the main methods include the following. One is a ray-based method, which primarily uses the time difference of arrival between the direct and sea-surface reflected arrivals to estimate the source depth; this method is mainly applicable to conventional deep-sea areas. Another method is based on normal modes, which utilizes the difference in excitation strength of different modes at different frequencies by the source depth to estimate the source depth; this method is mainly applicable to conventional shallow-sea areas.

In the deep Arctic Ocean, estimating the depth of an underwater source is more complicated than in conventional deep and shallow seas. Specifically, different methods are required for source depth estimation depending on the acoustic field characteristics under different conditions. For example, in the surface water column of the deep Arctic, the sound field consists of multiple modes and multipath rays; thus, modal or ray methods can be used for localization. In deeper water columns of the deep Arctic, the sound field is dominated by multipath rays; therefore, ray-based methods are primarily used for localization.

At present, methods for underwater source depth estimation in the Arctic mainly include the following: using synchronized vertical arrays for matched-field localization, using large-aperture vertical arrays for mode separation to achieve source range and depth estimation, or using the arrival elevation angle of deep refraction rays to estimate the source range, followed by the time difference of arrival between the source and sea-surface reflection to estimate the source depth. The above methods all require large-aperture synchronized vertical arrays, posing significant challenges for at-sea experiments. There are also source localization methods based on a single hydrophone, which involve complex mode separation to obtain the amplitude information of each mode, matching modal dispersion to estimate the source range, and matching modal amplitudes to estimate the source depth; these methods are operationally complex. For single-hydrophone methods, some utilize the upper frequency limit characteristic of modes formed by the vertical distribution of modal eigenfunctions in the deep Arctic surface layer to estimate the source depth. However, this approach has some problems: the accuracy is affected by ice attenuation, and the receiver depth must be smaller than the source depth, forcing the receiver to be placed at a shallower depth, which is difficult for practical deployment.

This paper mainly studies the modal dispersion interruption phenomenon in the surface normal mode waveguide of the deep Arctic Ocean, analyzes its formation mechanism and its relationship with source and receiver depths, and explores a method for estimating the source depth using the dispersion interruption frequency. The specific contents of each section are as follows: theoretical analysis of the modal dispersion interruption phenomenon, a source depth estimation method based on modal dispersion interruption, simulation studies of the dispersion interruption phenomenon and depth estimation, and finally, experimental data validation of the dispersion phenomenon and source depth estimation.

## II. THEORETICAL ANALYSIS OF THE MODAL DISPERSION INTERRUPTION PHENOMENON

For the underwater sound field in the deep Arctic, the acoustic field exhibits different characteristics at different depth ranges. Therefore, different source depth estimation methods need to be considered for different depth ranges. For a typical deep Arctic sound speed profile environment, as shown in Fig. 1, within the upper 400 m, simulation and analysis results indicate that the sound field in this surface range is composed of normal mode arrivals and ray arrivals; thus, mode or ray methods can be used to estimate the source depth in this depth range. For the ray method, when the source range is known, the time difference of arrival between the direct and sea-surface reflected arrivals can be used to estimate the source depth. For the normal mode method, when the source range is unknown, the variation of modal amplitudes can be used to estimate the source depth.

For a typical deep Arctic sound speed profile environment, simulation and analysis results show that the sound field at moderate and large depths is mainly composed of ray arrivals; therefore, ray methods are primarily used for source depth estimation in these depth ranges, specifically using the time difference of arrival between the direct and sea-surface reflected arrivals.

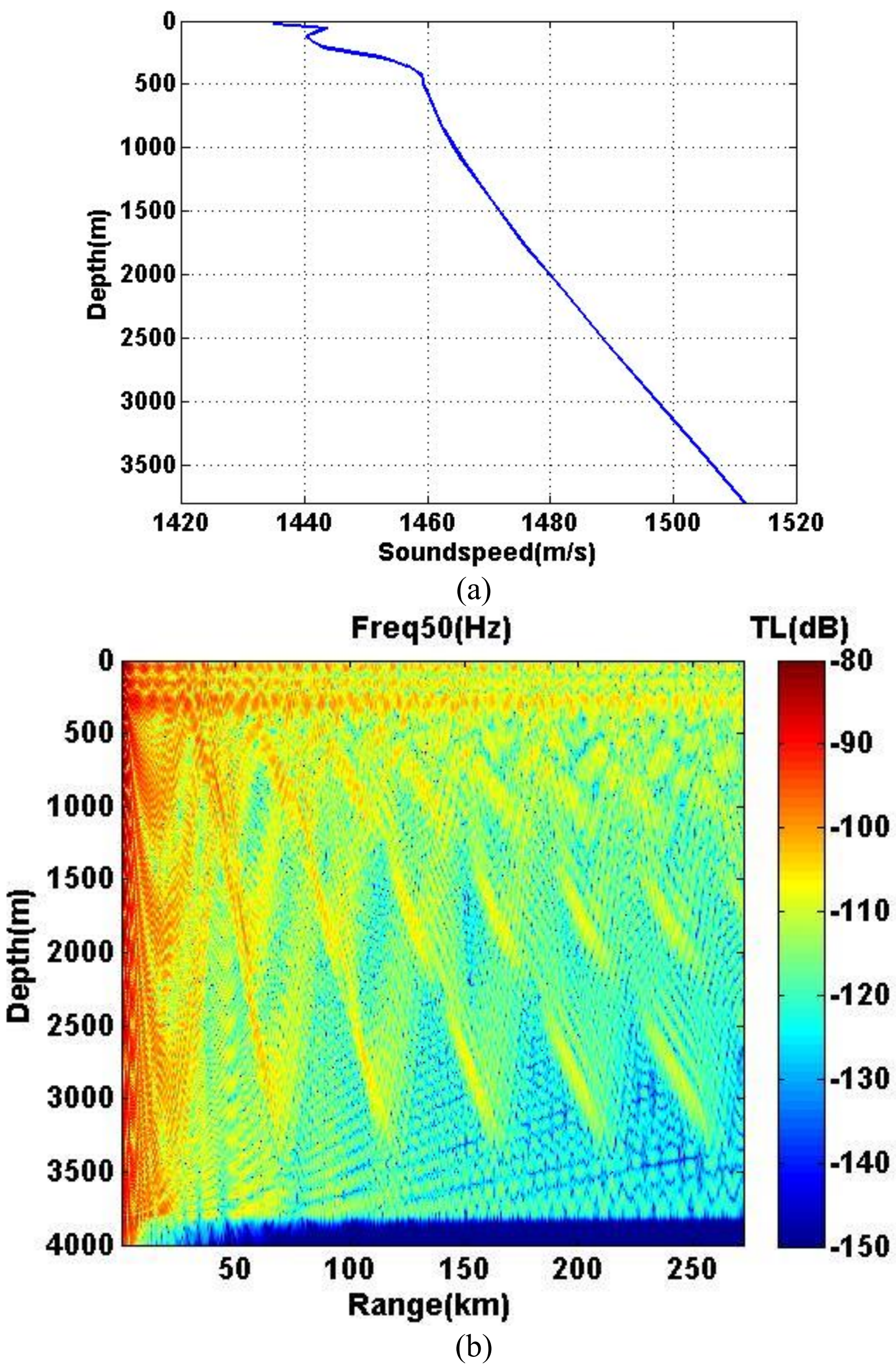

*Fig. 1. A typical deep Arctic sound speed profile and the typical deep-sea sound field under this condition: (a) sound speed profile, (b) spatial distribution of acoustic propagation loss.*

In the surface normal mode waveguide, the sound field propagates as a superposition of modes, as shown in Eq. (1):

$$P(\omega,r,z)=S(\omega)\frac{je^{-j\pi/4}}{\rho(z_s)\sqrt{8\pi r}}\sum_{m=1}^{M}\Psi_m(z_s)\Psi_m(z_r)\frac{e^{-\alpha_m(\omega)r}e^{jk_{rm}(\omega)r}}{\sqrt{k_{rm}(\omega)}} \tag{1}$$

where $\omega$ represents the angular frequency, $k_{rm}(\omega)$ represents the horizontal wavenumber, $\alpha_m(\omega)$ represents the attenuation coefficient, $\Psi_m(z)$ represents the eigenfunction, $z_s$ and $z_r$ represent the source and receiver depths, $r$ represents the source-receiver range, $M$ represents the number of effective normal modes, and $S(\omega)$ S represents the source spectrum level.

Regarding the eigenfunctions related to source and receiver depths in the received sound pressure expression, under the deep Arctic sound speed profile condition, the normal mode acoustic field model Kraken is used to calculate the vertical distribution of eigenfunctions for each mode, as shown in Fig. 2. It can be seen from Fig. 2 that different modes have different vertical distributions; the same source depth can excite different mode intensities, enabling multiple samplings of the source depth information. Moreover, the vertical coverage of modal eigenfunctions increases with the mode number. Lower-order modes concentrate their energy mainly in the surface layer, and thus are relatively less affected by seabed topography changes, which is beneficial for low-frequency long-range source localization. Therefore, this paper focuses on low-order modes.

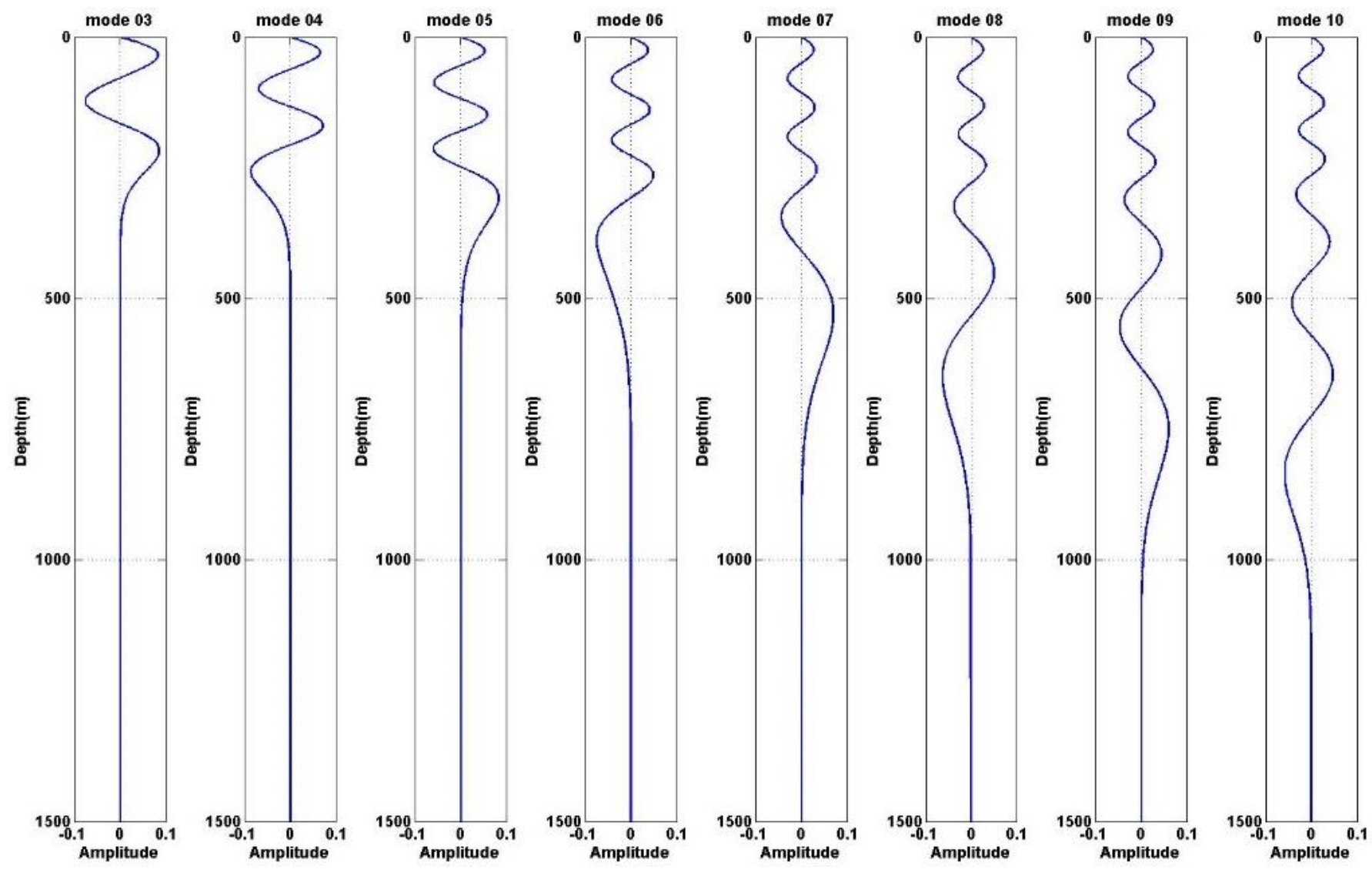

*Fig. 2. Eigenfunctions of modes 3 to 10 at 80 Hz.*

To explain the modal dispersion interruption phenomenon, i.e., the modal amplitude suddenly becomes approximately zero at a certain frequency, Fig. 3 shows the variation of modal eigenfunctions with frequency. When the source depth is exactly at a node of the eigenfunction of a certain mode at a certain frequency, the excitation amplitude becomes zero, forming an interruption in the dispersion curve. Taking Fig. 3 as an example, the first node depths from bottom to top for mode 7 at eight frequencies are 1428 m, 883 m, 610 m, 409 m, 300 m, 260 m, 239 m, and 225 m, respectively; for mode 9, the node depths at eight frequencies are 1768 m, 1113 m, 828 m, 633 m, 475 m, 346 m, 294 m, and 266 m. When the source depth is at any of these node depths, the modal amplitude is zero, manifesting as an interruption at that frequency in the dispersion structure. Therefore, this information can be used to infer the source depth: determine the dispersion interruption frequency of a mode in the observed data, use a normal mode propagation model to calculate the eigenfunction of that mode at that frequency, and the first node depth of this eigenfunction from bottom to top is the source depth. The first node is used because the amplitude of the mode below the first node is the largest, making the node effect more obvious in the time-frequency analysis of the acoustic signal.

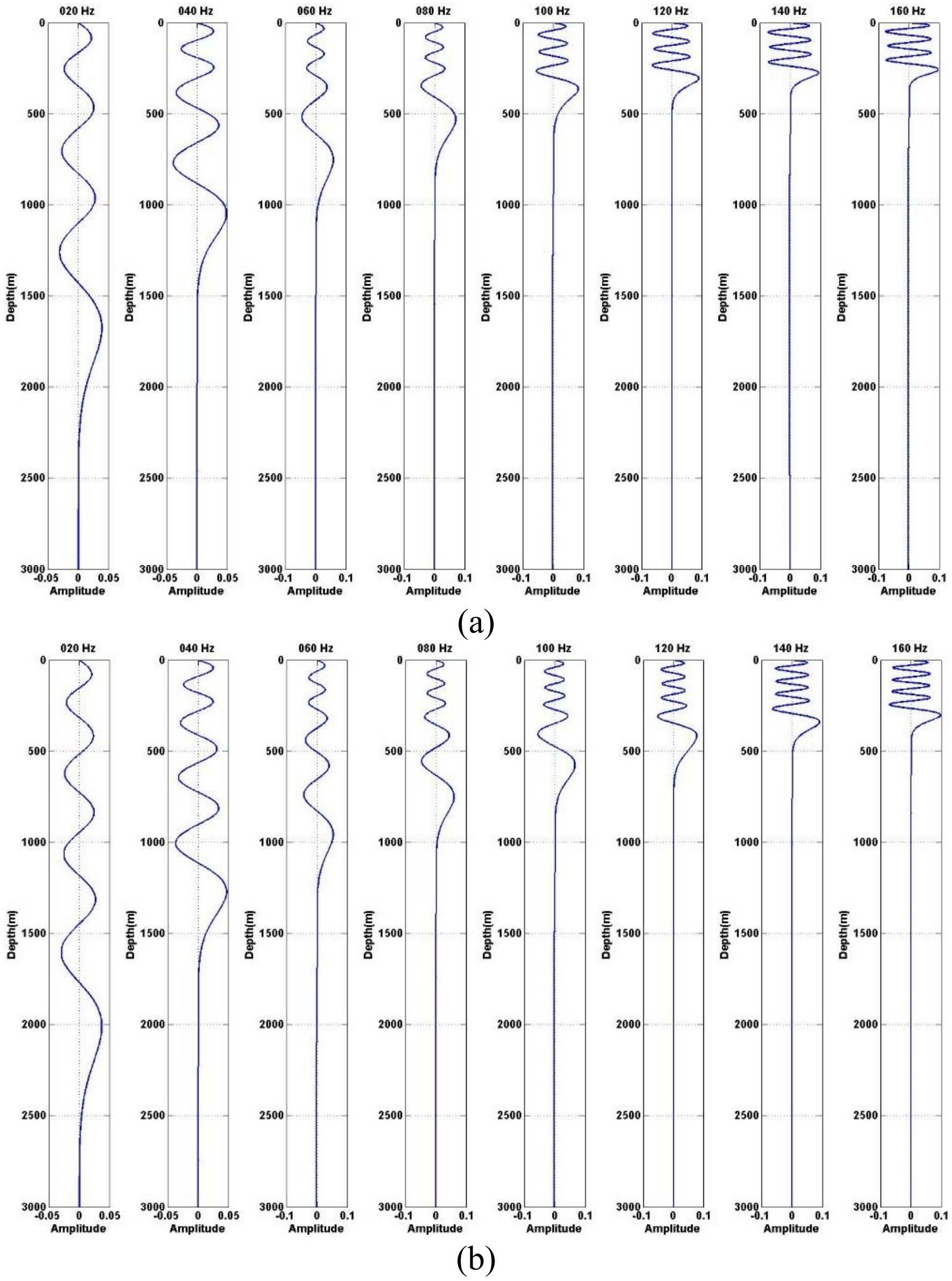

*Fig. 3. Eigenfunctions of mode 7 and mode 9 at different frequencies (20 Hz, 40 Hz, 60 Hz, 80 Hz, 100 Hz, 120 Hz, 140 Hz, and 160 Hz): (a) mode 7, (b) mode 9.*

## III. SOURCE DEPTH ESTIMATION BASED ON MODAL DISPERSION INTERRUPTION

Based on the normal mode characteristics of the surface sound field in the deep Arctic, when the source depth is at a node of a certain mode at a certain frequency, a dispersion interruption phenomenon appears in the dispersion structure of the received signal. Therefore, the dispersion interruption frequency of the received signal can be used to estimate the source depth.

If the dispersion interruption frequency of a mode is exactly in the middle of the dispersion curve, it can be clearly seen in the time-frequency analysis of the received signal. The premise is that the dispersion interruption frequency lies between the upper and lower frequency limits of the mode. The upper frequency limit of the mode is determined by the maximum of the source and receiver depths, while the lower frequency limit is affected by the seabed topography; therefore, comprehensive consideration is needed for the possibility of modal dispersion interruption. At the same time, due to sound propagation attenuation, only low-order modes in the received signal have sufficient signal-to-noise ratio; thus, this method can only be used for low-order modes. In addition, since this method requires a clear dispersion structure, it is applicable to low-frequency, long-range, broadband sources.

This method mainly considers three cases.

The first case is when both the receiver depth and source depth are within the surface 400 m range, and the receiver depth is significantly smaller than the source depth. In this case, there are multiple interruption bands in the dispersion structure, and the dispersion curve is divided into multiple segments. The multiple interruption bands correspond to multiple node frequencies of the source depth and receiver depth, respectively.

The second case is when both the receiver depth and source depth are within the surface 400 m range, and the receiver depth is close to the source depth. In this case, there is a wide interruption band in the dispersion structure, which is related to both the source depth and receiver depth.

The third case is when the receiver depth is in the range of 500–600 m, and the source depth is within the surface 400 m range. In this case, there is a narrow interruption band in the frequency structure, which is only related to the source depth.

Theoretically, the interruption frequency of a modal dispersion curve should correspond to the first node position of the eigenfunction from bottom to top, and the upper frequency limit corresponds to the maximum depth position of the modal eigenfunction. Therefore, the source depth can be estimated by simulating the first node depth of the mode from bottom to top at the interruption frequency. In addition, the modal dispersion interruption band should correspond to the first node positions of both the source depth and receiver depth, which can be solved if the receiver depth is known.

The following figure shows the depth of the first node of modes at different frequencies. It can be seen that the depth of the first node decreases as the frequency increases; therefore, the smaller the source or receiver depth, the higher the corresponding node frequency.

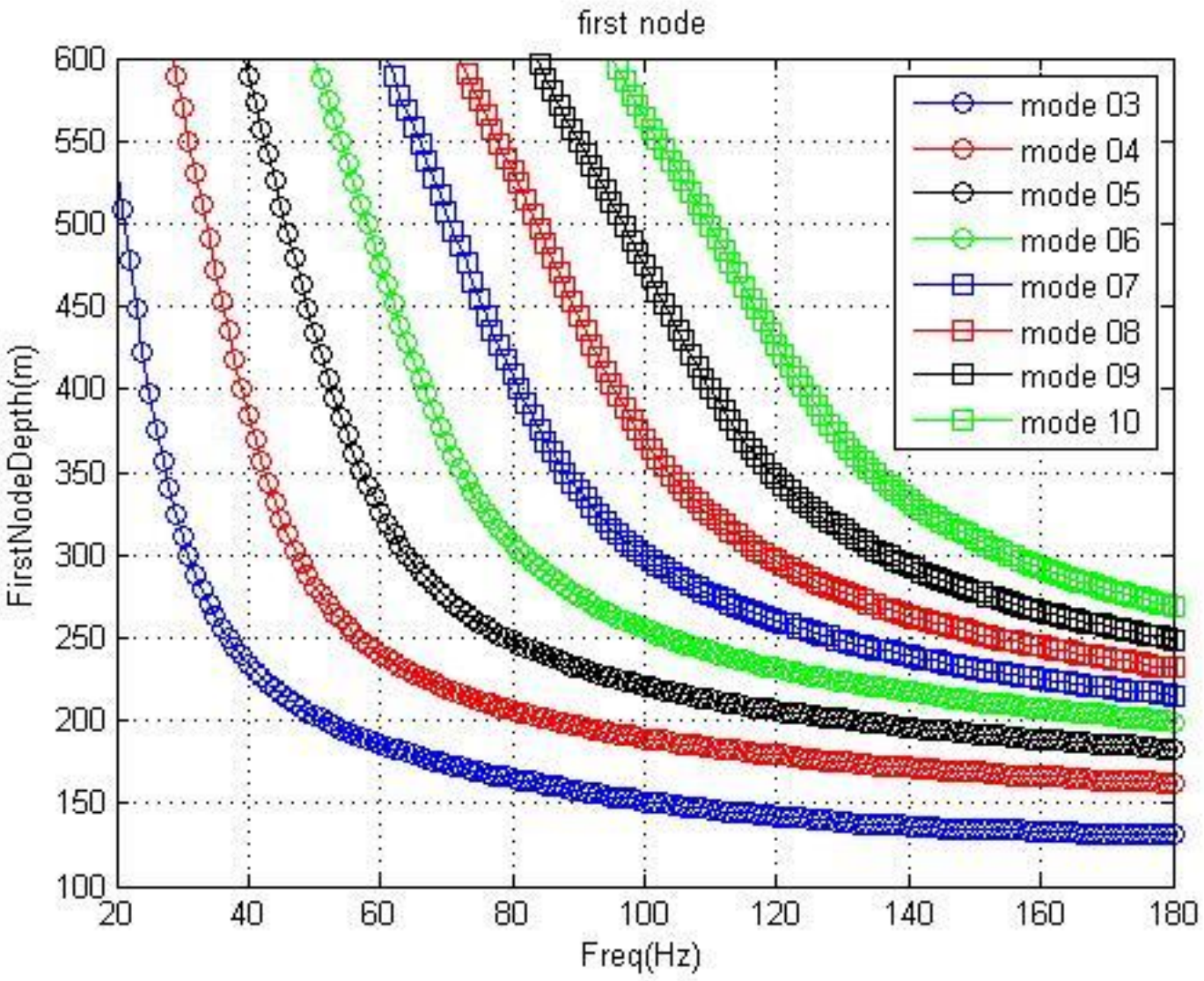

*Fig. 4. Curve of the depth of the first node of the modes as a function of frequency at different frequencies.*

The reason for using the first node is that, as the frequency decreases, the first node following the maximum amplitude has the highest signal-to-noise ratio and is the clearest; therefore, this paper uses the first node from bottom to top for source depth estimation. The first interruption after the strongest amplitude from high to low frequency corresponds to the first node of the modal eigenfunction from bottom to top.

The following figure shows the curves of the second, third, and fourth node depths varying with frequency for different modes. When the receiver depth is in the shallow part of the surface layer, multiple nodes need to be considered to exclude the multiple frequency interruptions caused by the receiver depth. Compared with Fig. 4, it can be seen that the node frequency decreases as the node number increases. That is, the same receiver depth has multiple nodes for the same mode, and the frequency of each node decreases as the node number increases. Taking a receiver depth of 250 m and mode 7 as an example, the four node frequencies are at 120 Hz, 90 Hz, 70 Hz, and 50 Hz, respectively.

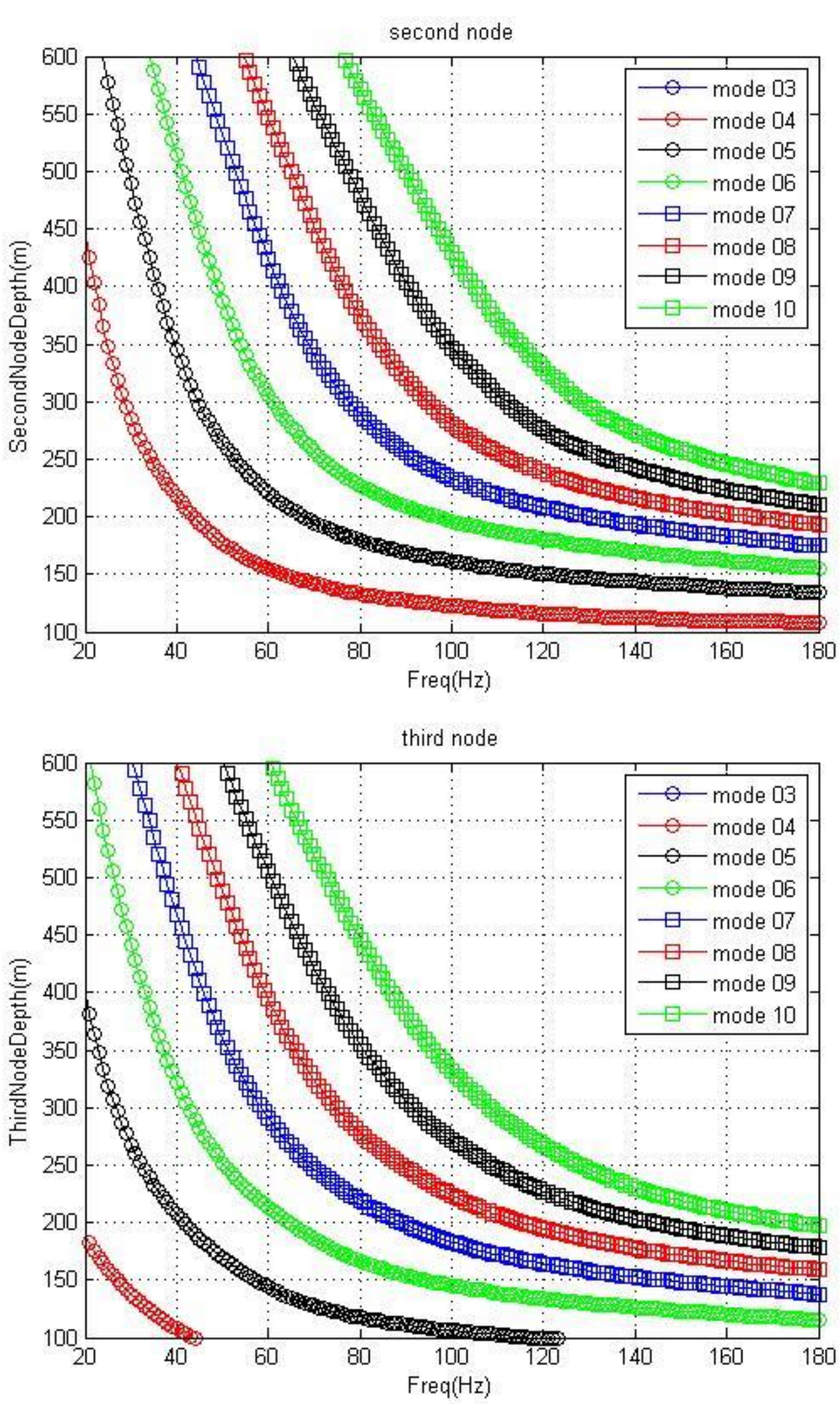
second node
600
550
500
450
400
350
300
250
200
150
100
SecondNodeDepth(m)
20
40
60
80
100
120
140
160
180
Freq(Hz)
mode 03
mode 04
mode 05
mode 06
mode 07
mode 08
mode 09
mode 10
third node
ThirdNodeDepth(m)
Freq(Hz)

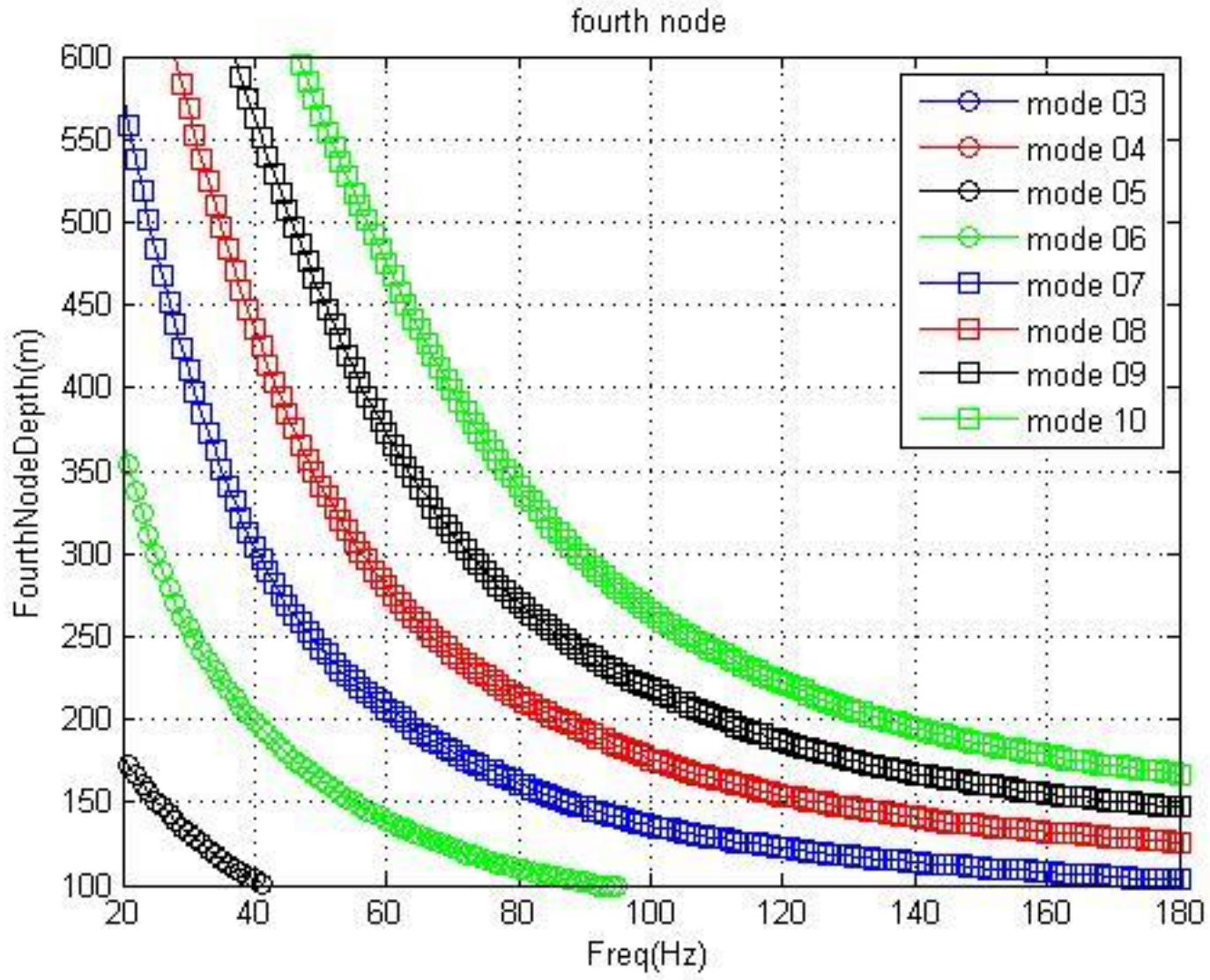

*Fig. 5. Curves of the depths of the second, third, and fourth nodes of the modes as a function of frequency.*

## IV. SIMULATION OF MODAL DISPERSION INTERRUPTION AND SOURCE DEPTH ESTIMATION

The following simulation analysis of acoustic signals is carried out for two cases: one where both the receiver depth and source depth are within the surface 400 m range, and another where the receiver depth is in the range of 500–600 m and the source depth is within the surface 400 m range, to analyze the effectiveness of the source depth estimation method at different receiver depths.

During the simulation, the source depth is set to 300 m, and the receiver depths are set to 134 m, 154 m, 174 m, 194 m, 224 m, 254 m, 284 m, 324 m, 494 m, and 534 m, totaling 10 receiver depths, divided into three groups: receiver depth less than source depth, receiver depth close to source depth, and receiver depth greater than source depth, to analyze the dispersion characteristics of the received signal at different receiver depth ranges. To compare with the experimental data, the simulation frequency band is set to 10 Hz to 160 Hz, and the propagation distance is consistent with the first station distance of the experiment (270.0128 km). The simulation uses a typical deep Arctic sound speed profile obtained from the experiment, with a sea depth of 3800 m. The normal mode acoustic field model Kraken is used to obtain the broadband complex sound pressure, which is then inversely Fourier transformed into a time-domain acoustic signal. Subsequently, time-frequency analysis is performed to obtain the modal dispersion structure, and the relationship between the modal dispersion interruption phenomenon and the source/receiver depths is analyzed.

## 1. Simulation of the modal dispersion interruption phenomenon

When the receiver depth is much smaller than the source depth, the time-frequency analyses of the received signals at different receiver depths are shown in Fig. 6. It can be seen that from mode 3 to mode 10, there is a clear dispersion interruption phenomenon, and different receiver depths have different dispersion interruption frequencies. The dispersion curve of each mode is divided into two or three segments. The high-frequency part has larger energy, corresponding to the first wave packet of the eigenfunction from bottom to top. The subsequent dispersion interruption frequency points are mainly due to the source depth or receiver depth being at the first node depth. Since the receiver depth is smaller than the source depth at this time, the interruption frequency caused by the receiver depth is higher than that caused by the source depth. Therefore, as the mode frequency decreases from high to low, two dispersion interruptions occur sequentially: first the interruption caused by the receiver depth, then the interruption caused by the source depth. For example, in Fig. 6(f), modes 5 to 9 clearly show a three-segment dispersion structure. In addition to the first node caused by the source and receiver in Fig. 6, there are other dispersion interruptions in higher frequency bands, which are the second or third nodes caused by the source depth. In actual source depth estimation, it is necessary to use the known receiver depth to exclude all node depths caused by the receiver in advance, so as to obtain the first node frequency caused by the source depth, and then accurately estimate the source depth. Therefore, the source depth estimation method in this paper is more suitable for cases where the source depth is relatively large.

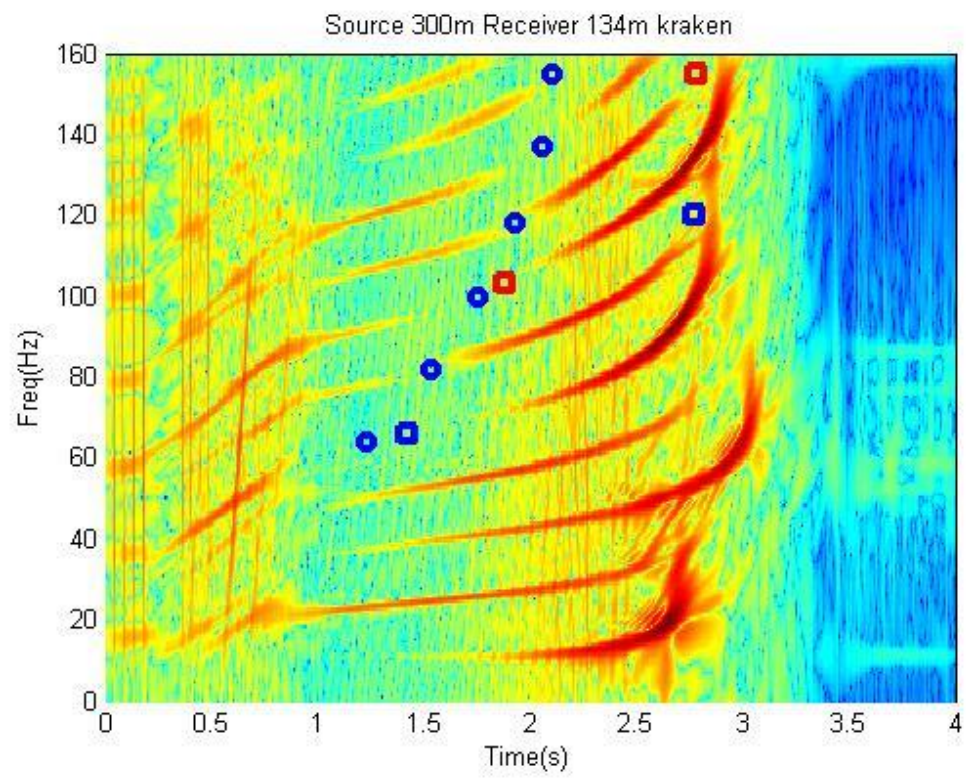

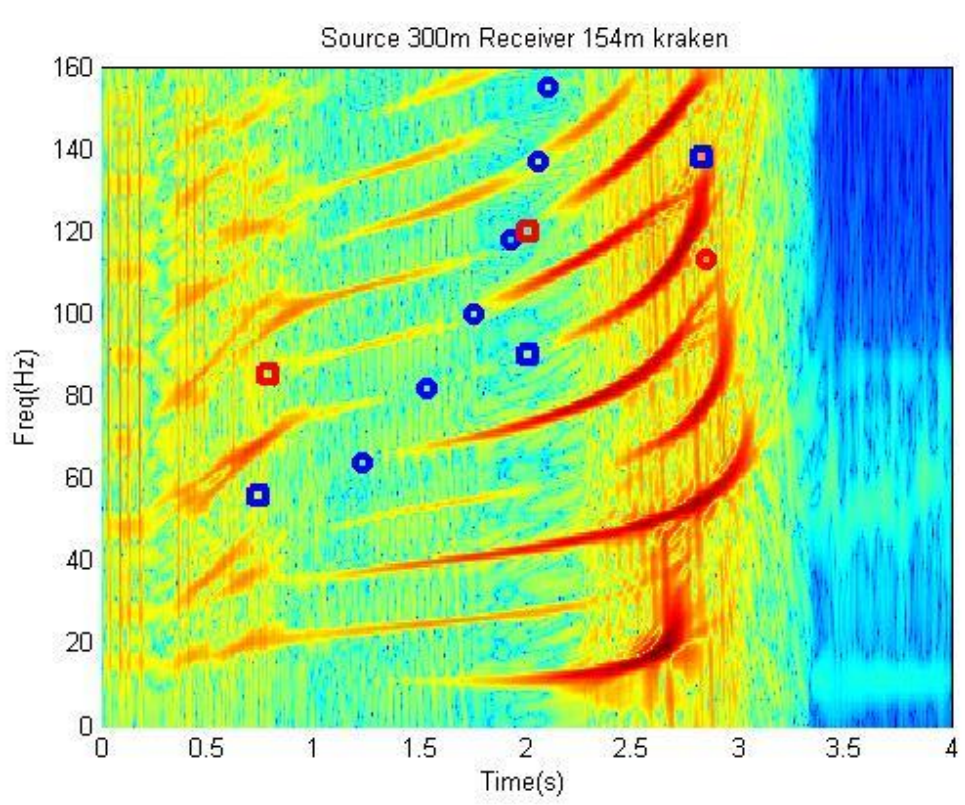

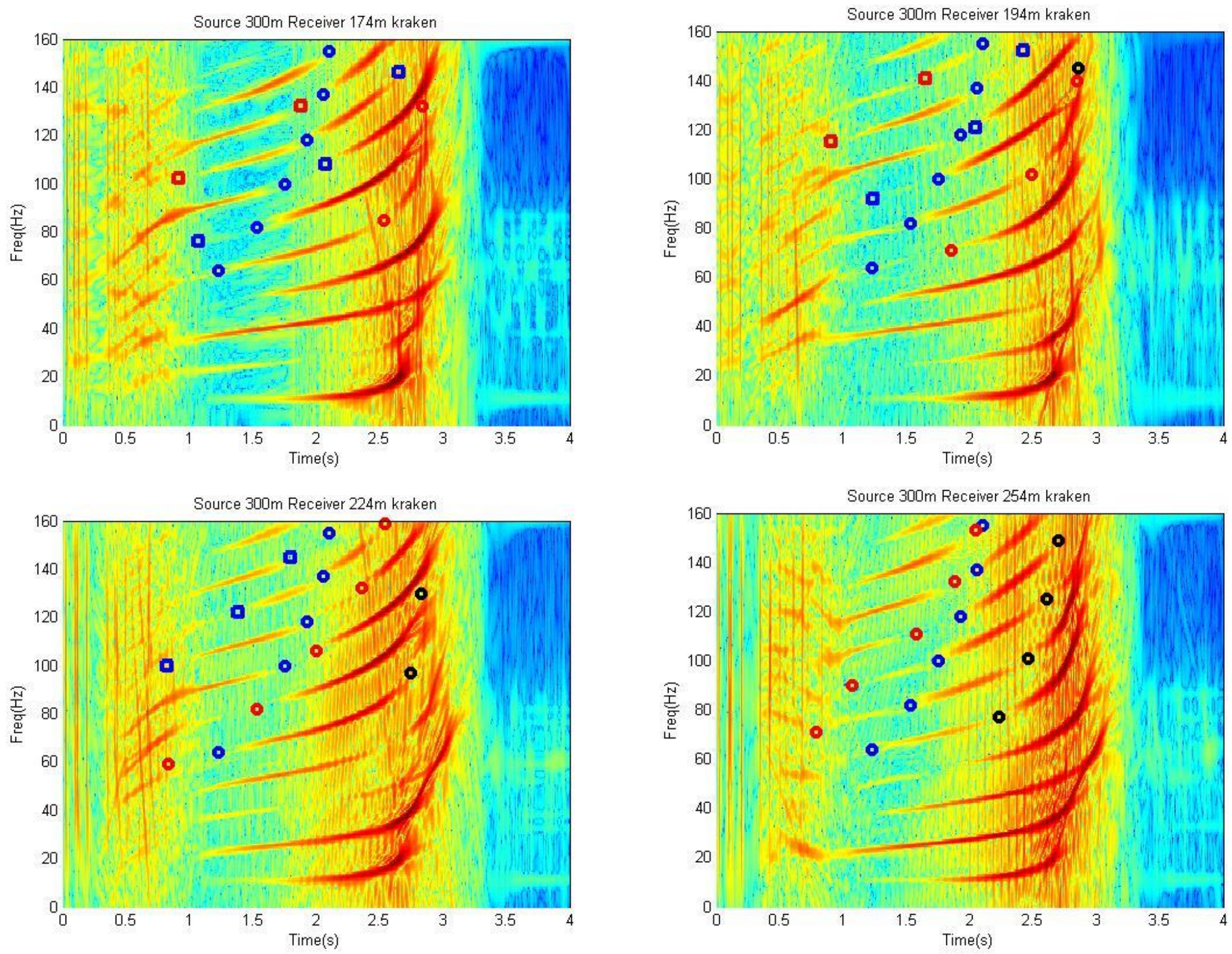

*Fig. 6. Dispersion structures of the acoustic signals at six depths (134 m, 154 m, 174 m, 194 m, 224 m, and 254 m), and the dispersion interruption positions estimated from the source depth and receiver depth (blue circles: first-node dispersion interruption frequencies corresponding to the source depth; black circles: first-node dispersion interruption frequencies corresponding to the receiver depth; red circles: second-node dispersion interruption frequencies corresponding to the receiver depth; blue squares: third-node dispersion interruption frequencies corresponding to the receiver depth; red squares: fourth-node dispersion interruption frequencies corresponding to the receiver depth).*

When the receiver depth is close to the source depth, the time-frequency analyses of the received signals at different receiver depths are shown in Fig. 7. From Fig. 7, it can be seen that modes 5 to 10 have regular dispersion interruption phenomena, and the interrupted parts have a wide frequency band. The reason for forming a regular wide band is that the receiver and source depths are close, causing the amplitude to be small in a frequency band, forming an interruption band, which will be analyzed in detail later. From Fig. 6, it can also be seen that when the receiver depth is close to the source depth, the dispersion interruption phenomenon is relatively regular, which is beneficial for quickly estimating the source depth.

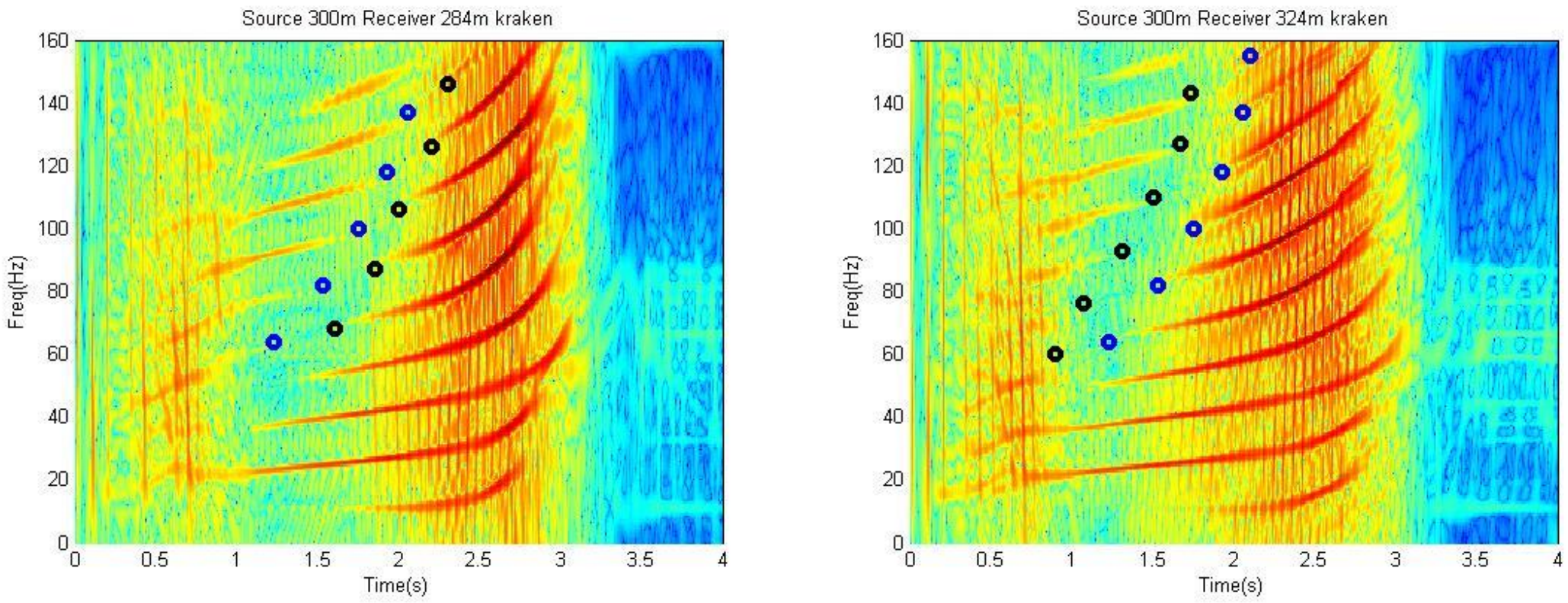

*Fig. 7. Dispersion structures of the acoustic signals at two receiver depths (284 m and 324 m), and the dispersion interruption positions estimated from the source depth and receiver depth (blue circles: first-node dispersion interruption frequencies corresponding to the source depth; black circles: first-node dispersion interruption frequencies corresponding to the receiver depth).*

When the receiver depth is much larger than the source depth, the time-frequency analyses of the received signals at different receiver depths are shown in Fig. 8. The dispersion interruption phenomenon at larger depths is different from the previous two cases. From mode 6, the high-frequency part basically disappears, because the receiver depth is large, causing the high-frequency part to be unexcitable. For modes 1 to 5, the high-frequency part can still be excited, because the eigenfunction coverage of lower-order modes is larger. Therefore, the dispersion interruption phenomenon can only be observed in the middle part, such as in mode 5.

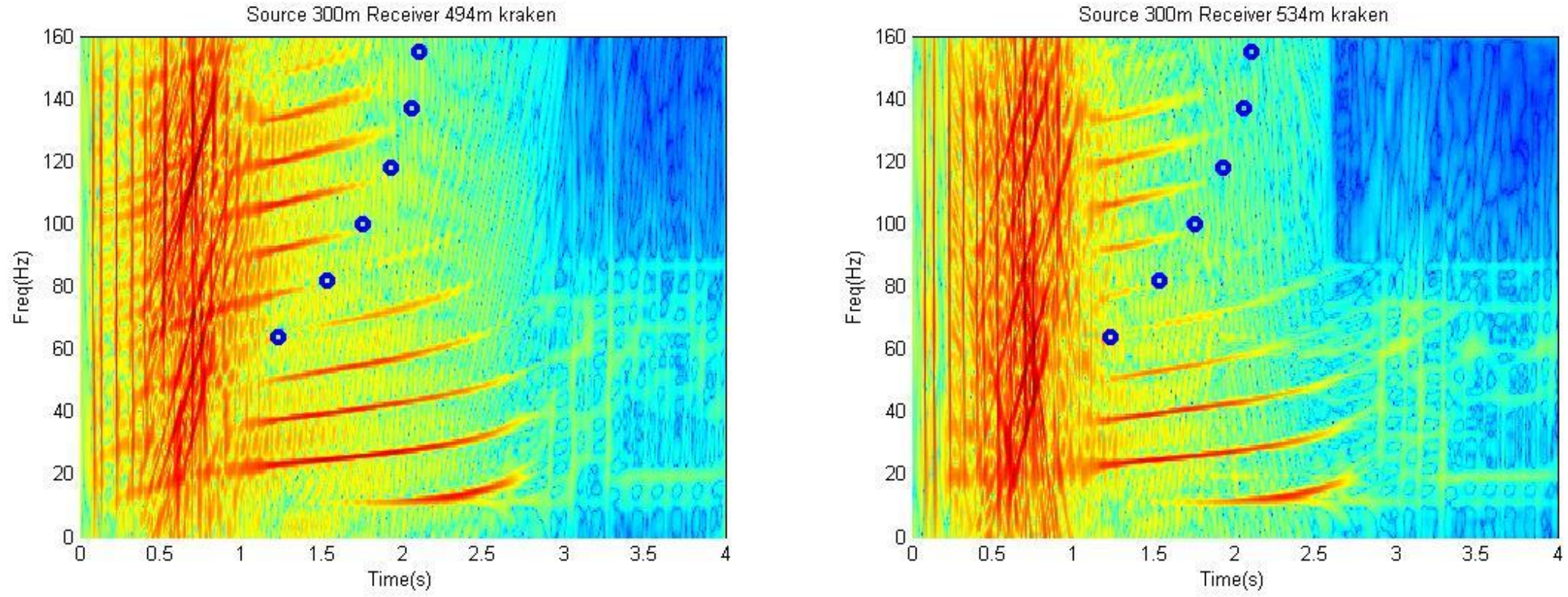

*Fig. 8. Dispersion structures of the acoustic signals at two receiver depths (494 m and 534 m), and the dispersion interruption positions estimated from the source depth (blue circles: first-node dispersion interruption frequencies corresponding to the source depth).*

## 2. Modal dispersion structure with known source and receiver depths

The following uses an example at a single depth to explain in detail the formation of dispersion interruption. Taking mode 7 as an example, with fixed source (300 m) and receiver (284 m) depths, Fig. 9 shows the modal amplitude variation with frequency for the source and receiver depths. The right figure shows the product of the two, because the received signal is the result of the combined effect of source and receiver depths. It can be seen that because the source and receiver depths are close, their modal amplitude variation patterns are similar. The high-frequency part has the maximum amplitude, which can be seen in the modal dispersion structure simulation as the maximum value of the largest wave peak of the modal eigenfunction depth. Immediately after the maximum amplitude, there follows an interruption band formed by the first nodes corresponding to the source and receiver. We set the maximum value in the interruption band as the threshold, and frequencies on both sides lower than this threshold are identified as interruption frequencies.

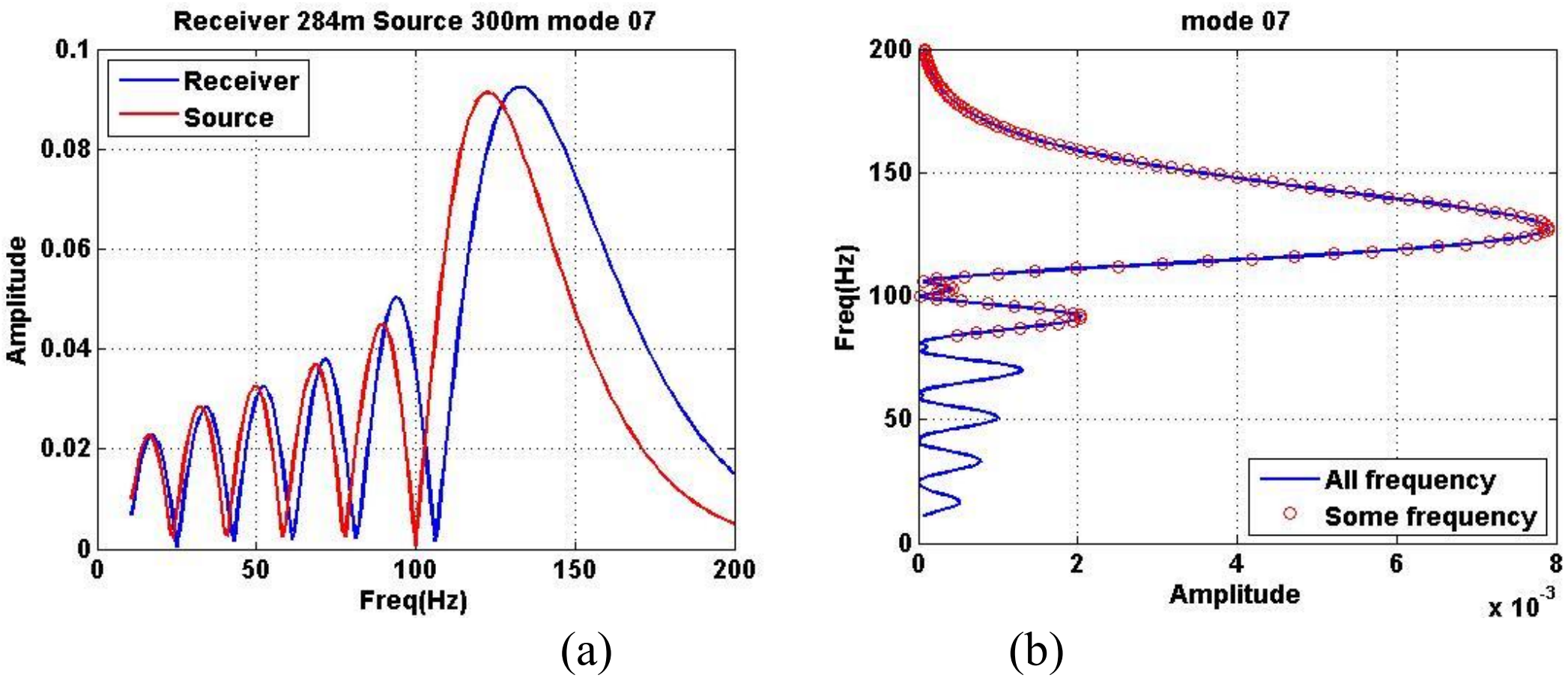

*Fig. 9. Modal amplitudes of mode 7 at different frequencies for the receiver depth and source depth: (a) amplitudes under the individual conditions, (b) amplitude under the combined condition.*

Similar calculations are performed for modes 5 to 10, as shown in Fig. 10. It can be seen that when the source depth and receiver depth are close, different modes have similar amplitude variation patterns with frequency, so under these conditions, it is beneficial to estimate the source depth.

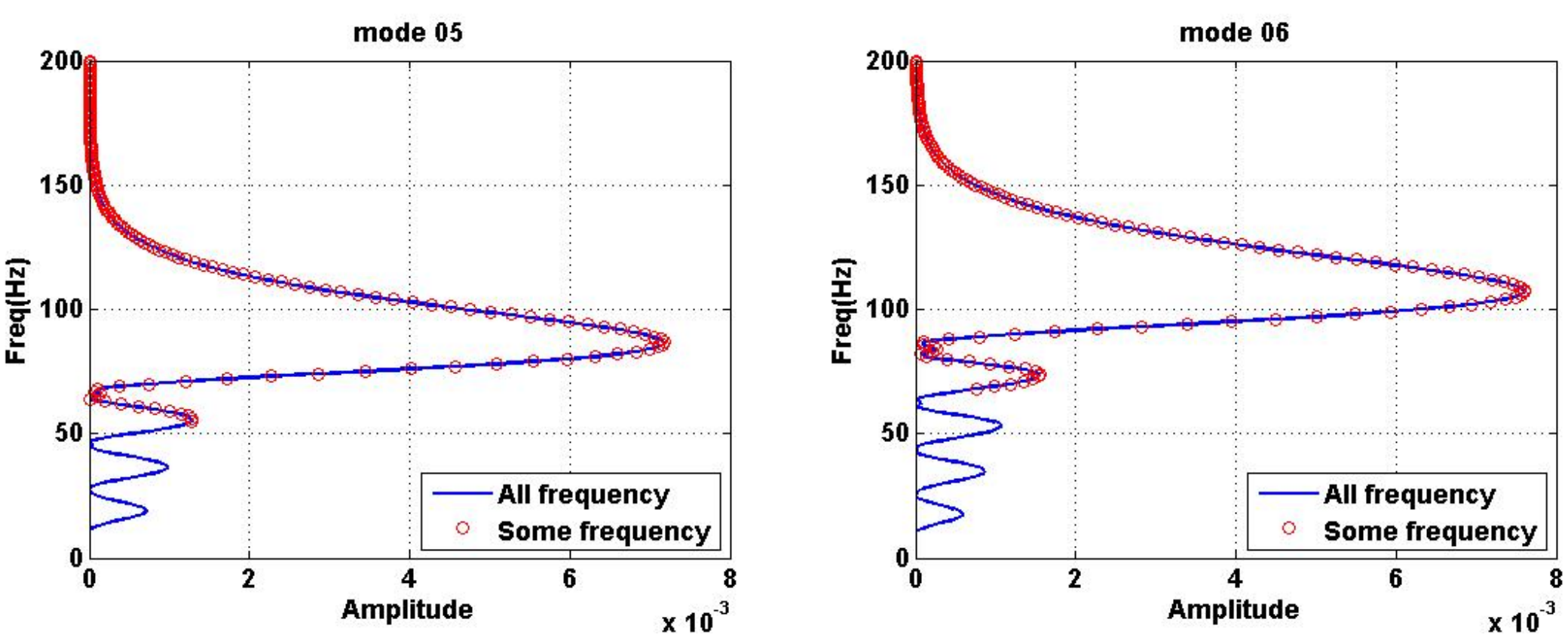

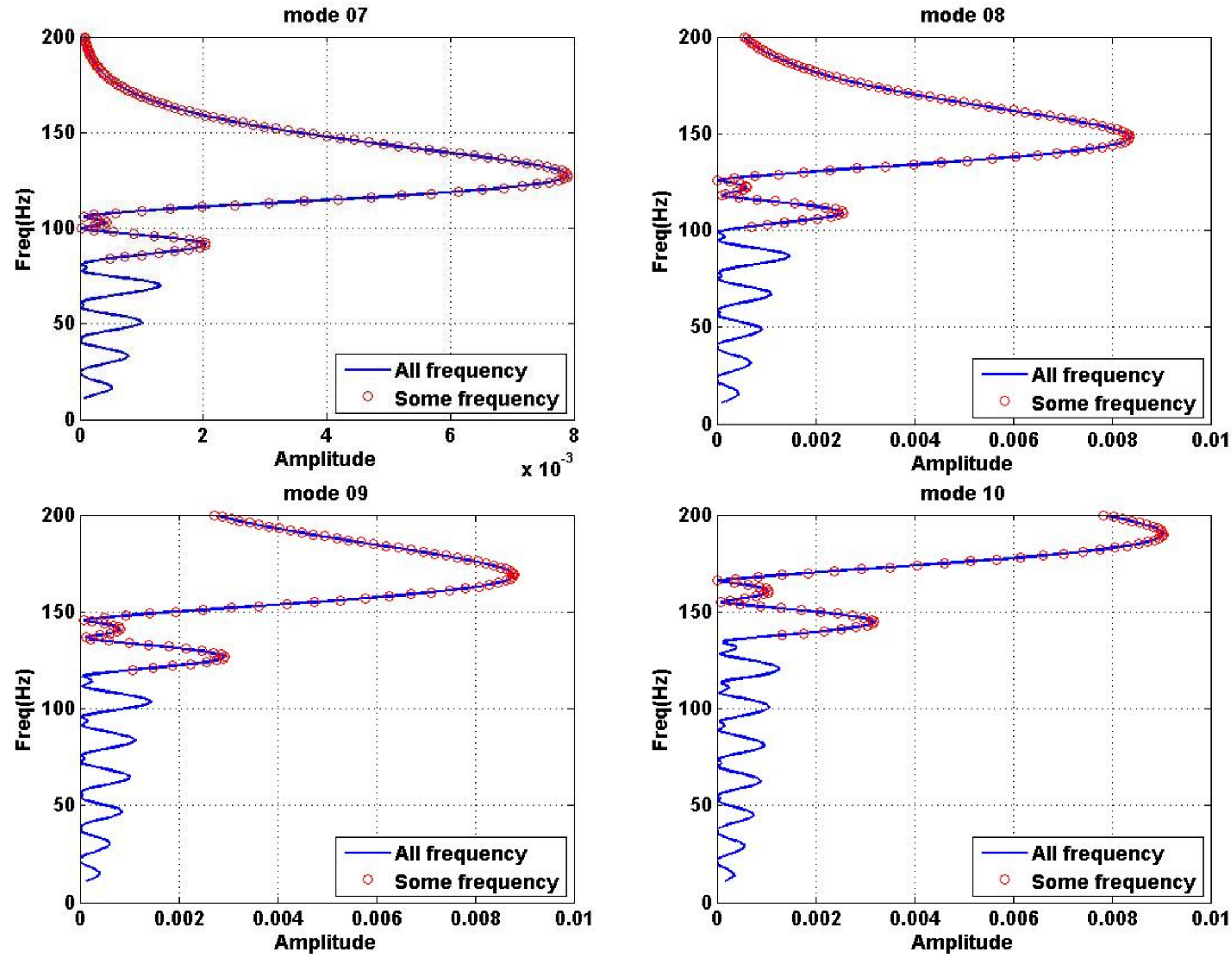

*Fig. 10. Modal amplitudes at different frequencies for modes 5 to 10 for the receiver depth and source depth.*

To further illustrate the dispersion interruption phenomenon, Fig. 11 gives the calculated modal group velocity curves. Then, the dispersion frequencies obtained from Fig. 10 are used to calculate the arrival times using the group velocity, and the obtained dispersion curves are compared with the simulated acoustic signals.

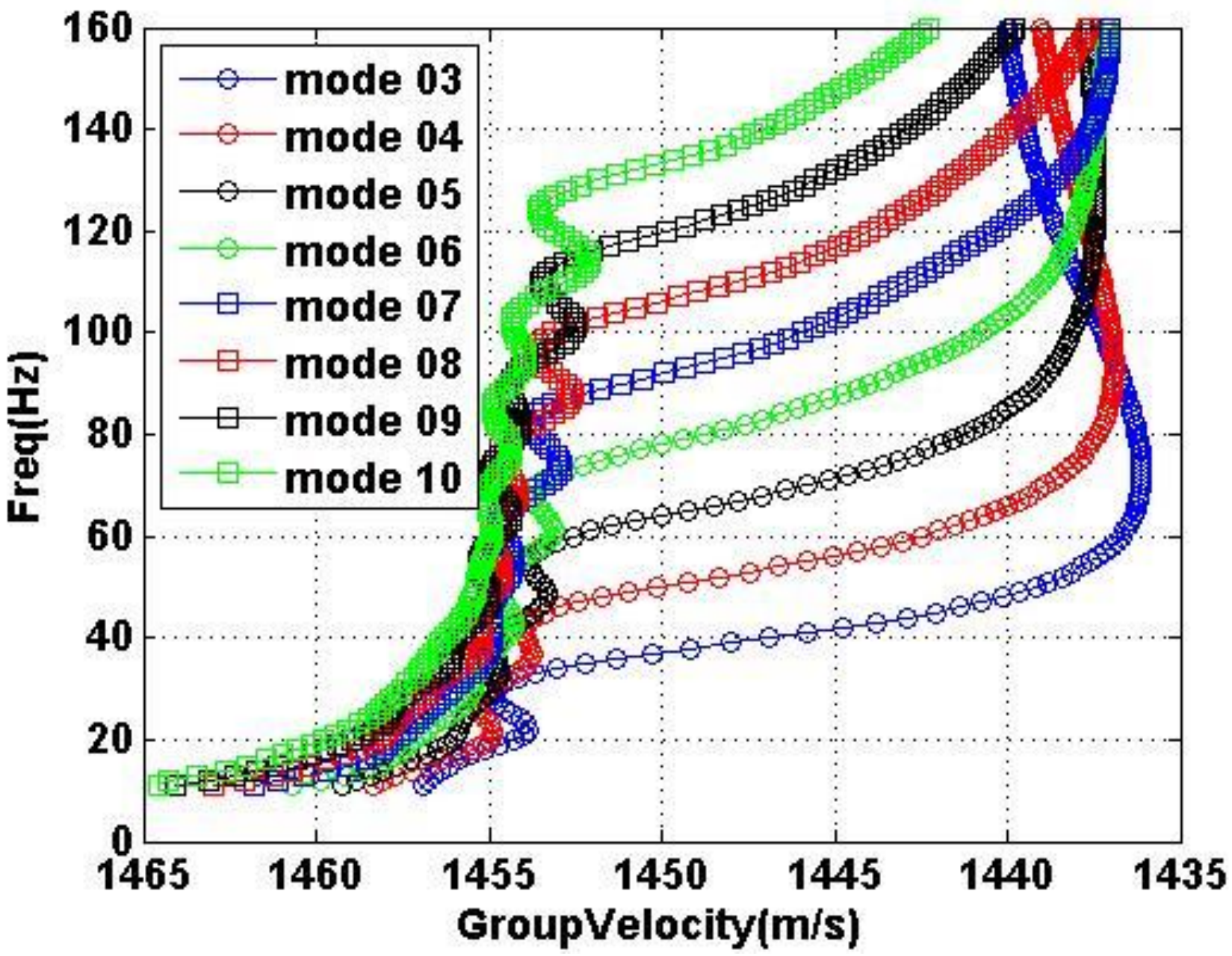

*Fig. 11. Calculated modal group velocity curves.*

Fig. 12 shows the comparison between the complete dispersion curves and the simulated signal time-frequency analysis, and also the comparison between the estimated dispersion interruptions and the simulated signal time-frequency analysis. It can be seen that the predicted dispersion interruptions are consistent with the simulation results, especially for the dispersion interruption frequencies of different modes. The estimated interruption frequency parts are completely consistent with the simulated dispersion results. The above simulation results well explain the reason for the modal dispersion interruption phenomenon in the deep Arctic surface waveguide, and show that when the receiver and source depths are close, the dispersion interruption is regular.

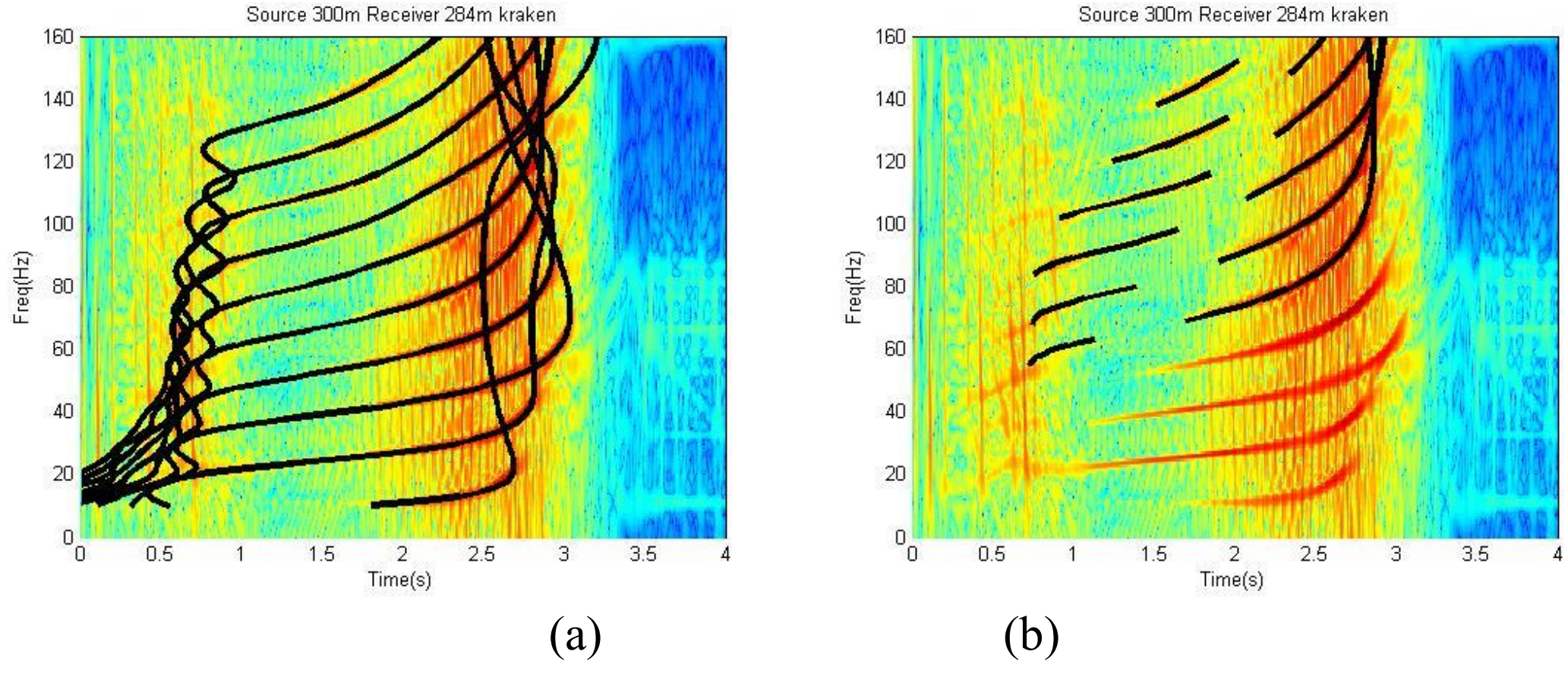

(a) (b)

*Fig. 12. Comparison between the predicted dispersion curves and the time-frequency analysis of the simulated*

*signal: (a) complete dispersion curves, (b) interrupted dispersion curves.*

## V. EXPERIMENT ON MODAL DISPERSION INTERRUPTION AND SOURCE DEPTH ESTIMATION

Next, acoustic propagation data from an Arctic experiment are used for experimental validation. A broadband pulse source propagation experiment conducted in the Chukchi Plateau and the central ice zone in 2025 is adopted. To achieve long-distance propagation of low-frequency signals and obtain a clear dispersion structure in the received signal, a high-source-level broadband pulse source was used, with the source level in the low-frequency band exceeding 200 dB. To test the source depth estimation effect at different receiver depths, multiple self-contained hydrophone receivers at different depths were used to form a mooring system. This system had 14 hydrophones within a depth range of 600 m, designed to be at depths of 080 m, 110 m, 130 m, 150 m, 170 m, 200 m, 230 m, 260 m, 300 m, 350 m, 420 m, 470 m, 510 m, and 590 m. Each hydrophone was equipped with a temperature and pressure recorder to collect real-time working depth information of the receiver.

The experimental sea area is in the Chukchi Plateau and the central ice zone. The receiving mooring was deployed in the Chukchi Plateau, and the source station was located in the deep-sea area near the Chukchi Plateau, at a distance of about 272 km from the receiving station. Therefore, the acoustic propagation in this experimental sea area is complex, including three-dimensional variations of seabed topography, changes in sea-ice cover, and three-dimensional variations of the sound speed profile. The following figure shows the seabed topography along the propagation path obtained from the seabed topography database, and the two-dimensional acoustic propagation loss simulation based on the source depth. It can be seen that the seabed topography is complex, changing from a deepest point of over 3000 m to a shallowest point of about 600 m seamount, and finally reaching the receiving station at a depth of about 2000 m. The seabed topography blocks the propagation of deep refraction sources, but its impact on normal mode propagation within the surface 600 m is reduced.

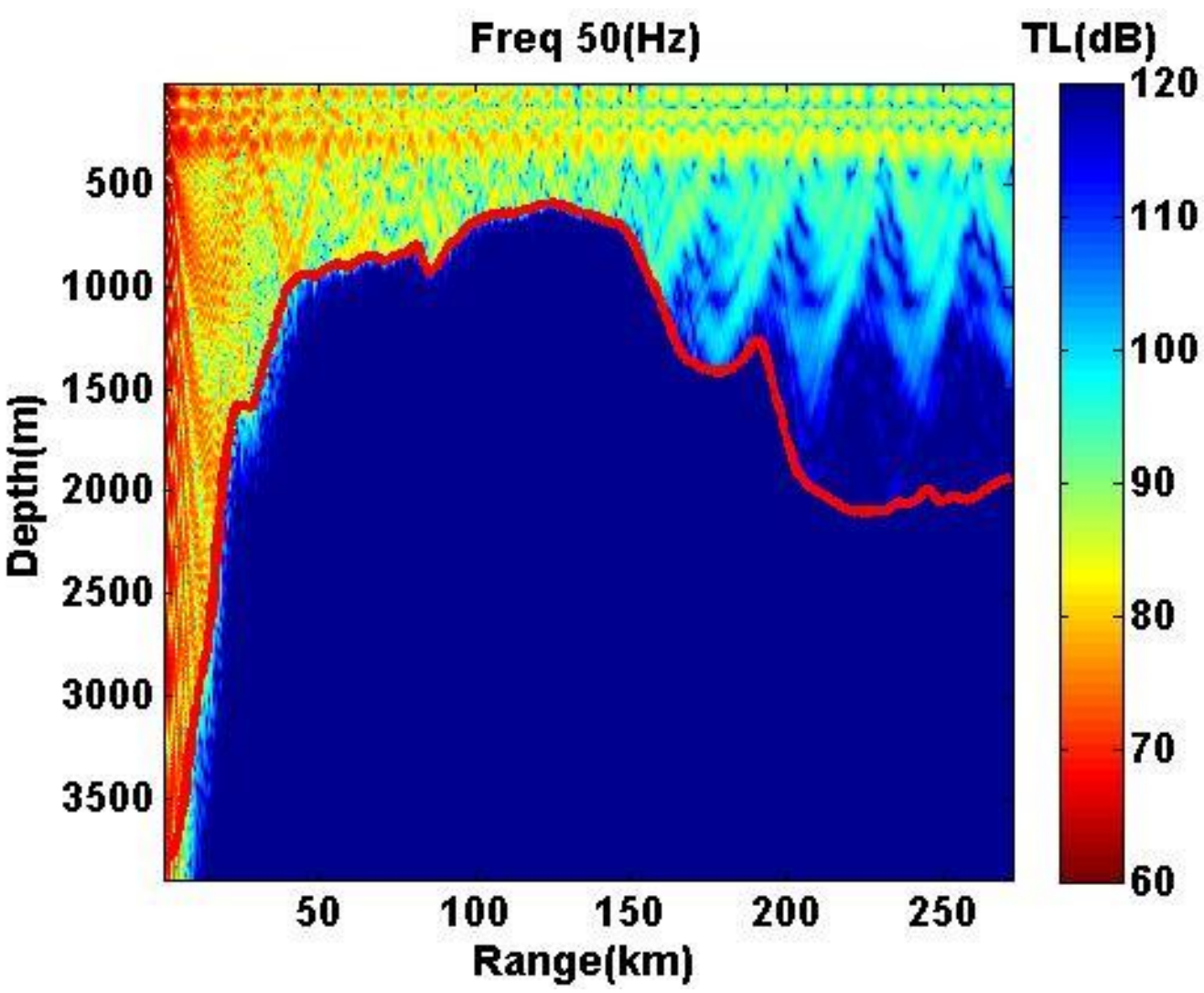

*Fig. 13. Two-dimensional acoustic propagation loss simulation based on the topography of the experimental sea area.*

In the experimental data, ten hydrophones with design depths of 110 m (HTD02), 130 m (HTD03), 150 m (HTD04), 170 m (HTD22), 200 m (HTD05), 230 m (HTD21), 260 m (HTD06), 300 m (HTD07), 470 m (HTD10), and 510 m (HTD11) are used for analysis. The calibrated source depth is 300 m. The first six hydrophones have depths less than the source depth, and the last two have depths significantly greater than the source depth. According to the depth recorder data analyzed later, the actual depths of the ten hydrophones are 121 m, 145 m, 164 m, 183 m, 212 m, 246 m, 276 m, 317 m, 485 m, and 531 m.

The following figure shows the time-frequency analysis of the acoustic signals from six channels in the deep Arctic surface layer, and the estimated modal dispersion interruption frequencies based on the source depth and receiver depth. It can be seen that when the receiver depth is shallow, in addition to the dispersion interruption caused by the source depth, there are multiple dispersion interruptions caused by the receiver depth, including the first, second, third, and fourth nodes. The estimated dispersion interruption frequency points are in good agreement with the experimental received signals, with some differences in time. This paper focuses on frequency; the arrival time is determined by the group velocity, which is determined by the sound speed profile. Since there is no sound speed profile along the propagation path, there is an error in the modal arrival time, especially for the second, third, and fourth node frequencies, which are affected by the complex variation of the surface sound speed.

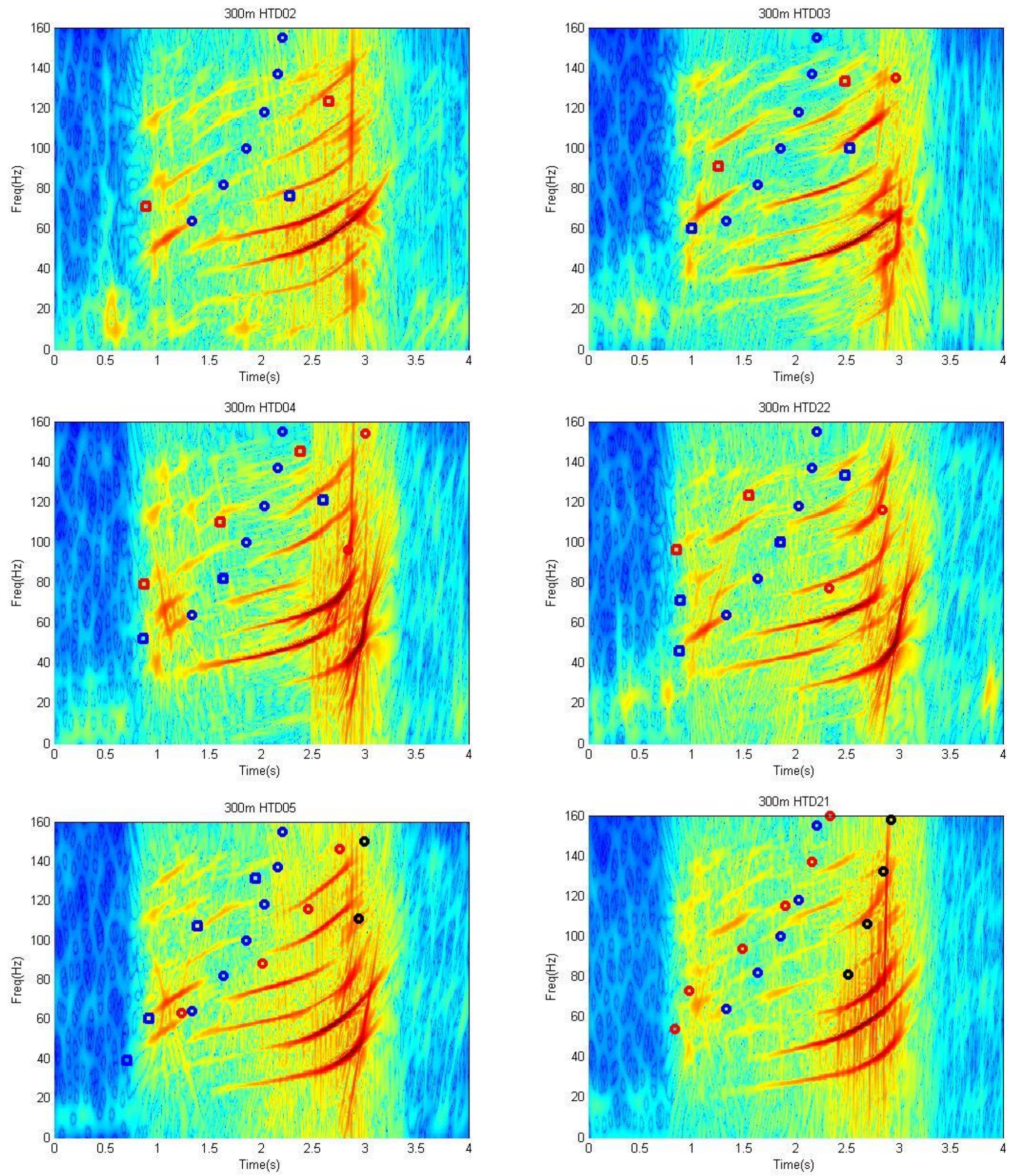

*Fig. 14. Six channels (HTD02 (110 m), HTD03 (130 m), HTD04 (150 m), HTD22 (170 m), HTD05 (200 m), and HTD21 (230 m)) at a propagation distance of 270.0128 km, and the dispersion interruption positions estimated from the source depth and receiver depth (blue circles: first-node dispersion interruption frequencies corresponding to the source depth; black circles: first-node dispersion interruption frequencies corresponding to the receiver depth; red circles: second-node dispersion interruption frequencies corresponding to the receiver depth; blue squares: third-node dispersion interruption frequencies corresponding to the receiver depth; red squares: fourth-node dispersion interruption frequencies corresponding to the receiver depth).*

From the comparison of HTD06 and HTD07 at the first station, it can be seen that the greater the difference between the receiver depth and the source depth, the wider the modal dispersion interruption band. The dispersion interruption of HTD07 is much smaller than that of HTD06, because the depth of HTD07 is closer to the source depth. Whether the upper and lower limits of the interruption frequency are determined by the source depth and receiver depth, respectively: when the receiver depth is less than the source depth, the upper limit of the interruption frequency is determined by the receiver depth, and the lower limit is determined by the source depth; the larger the difference between the two depths, the wider the interruption band. When the receiver depth is greater than the source depth, the upper limit of the interruption frequency is determined by the source depth, and the lower limit is determined by the receiver depth. There is some difference in time between the estimated dispersion interruption frequency points and the experiment, but the article focuses on frequency; the arrival time is determined by the group velocity, which is determined by the sound speed profile. Since there is no sound speed profile along the propagation path, there is an error in the modal arrival time.

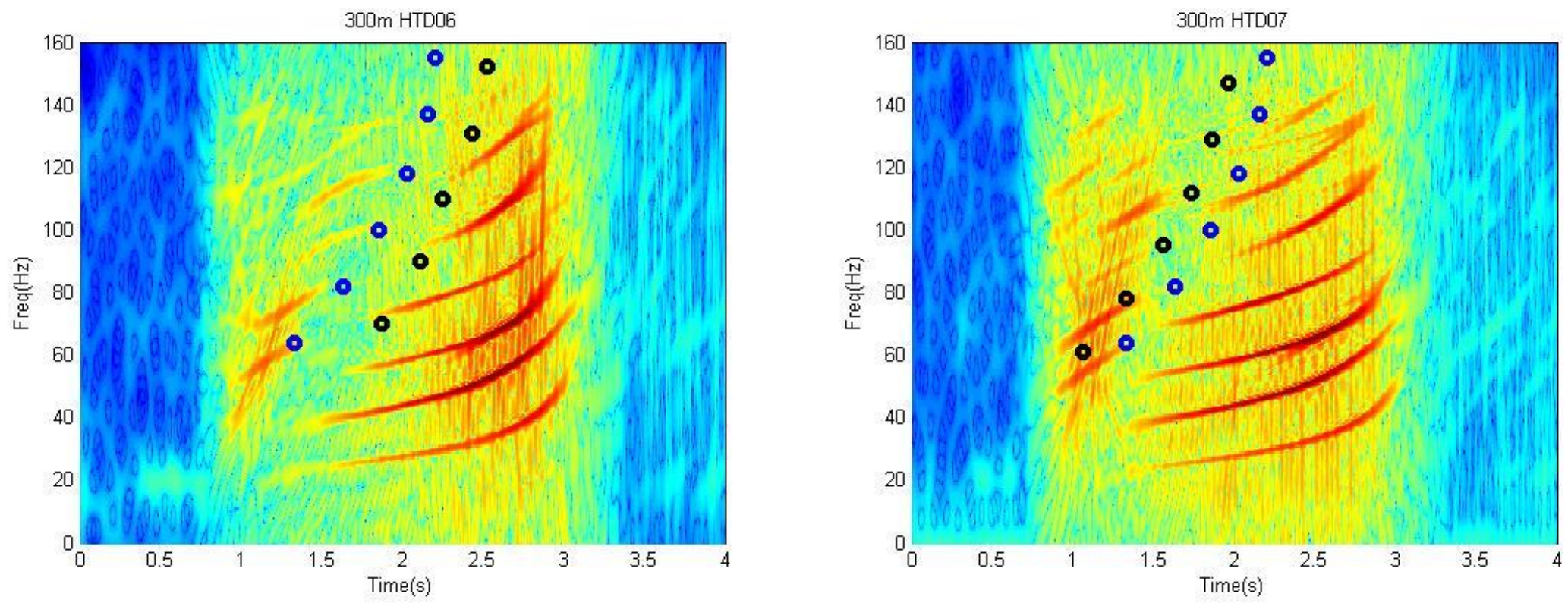

*Fig. 15. HTD06 (260 m) and HTD07 (300 m) at a propagation distance of 270.0128 km, and the dispersion interruption positions estimated from the source depth and receiver depth (blue circles: first-node dispersion interruption frequencies corresponding to the source depth; black circles: first-node dispersion interruption frequencies corresponding to the receiver depth).*

From the time-frequency analysis of the received signals at large depths in the following figure, it can be seen that the modal dispersion interruption phenomenon is mainly concentrated in mode 5, which is consistent with the simulation results. At large depths, the high-frequency part of higher-order modes has low energy. The interruption frequency of mode 5 estimated from the source depth is in good agreement with the experimental interruption frequency.

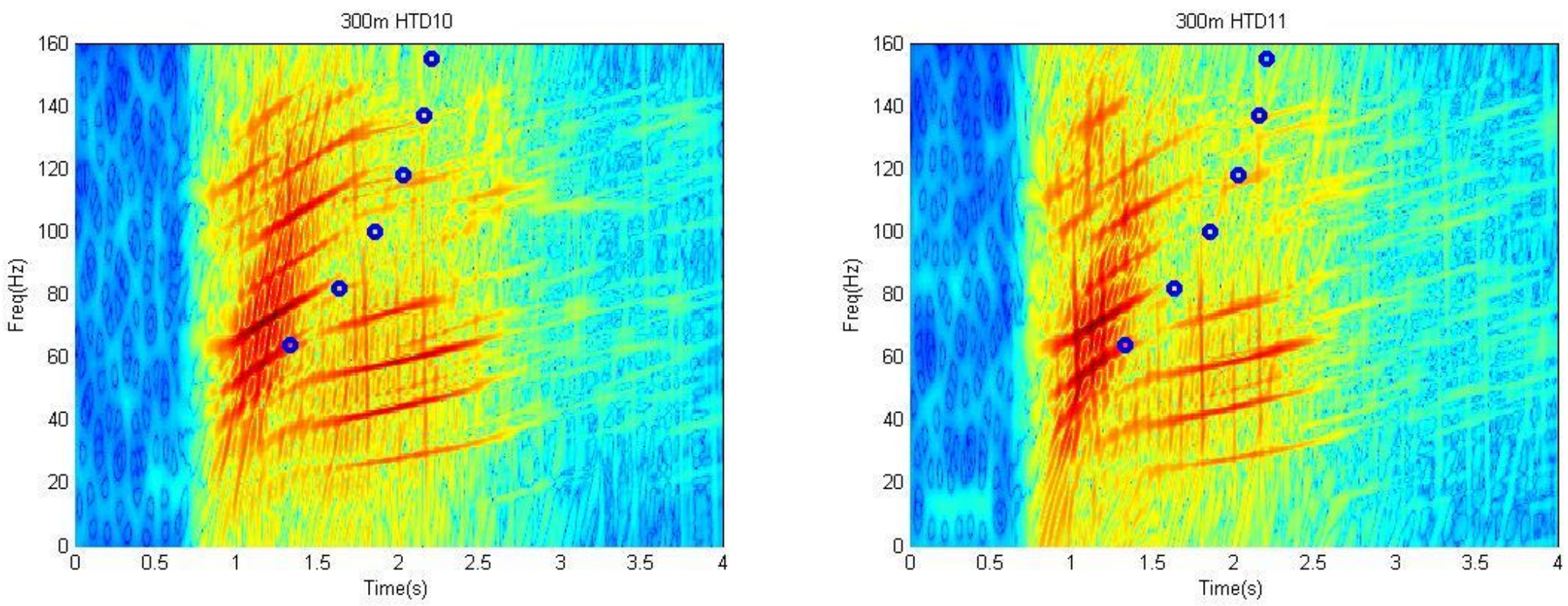

*Fig. 16. HTD10 (470 m) and HTD11 (510 m) at a propagation distance of 270.0128 km, and the dispersion interruption positions estimated from the source depth (blue circles: first-node dispersion interruption frequencies corresponding to the source depth).*

For the received signals at large depths, the frequency of the modal dispersion interruption part is only affected by the source depth; therefore, the source depth can be quickly estimated based on this interruption frequency, as shown above. This is the advantage of large-depth reception, but the receiver depth should not be too large, otherwise it will turn into ray reception with no dispersion effect.

# VI. CONCLUSIONS

For the surface acoustic waveguide in the deep Arctic Ocean, due to the unique sound speed profile, a normal mode waveguide is formed within approximately the upper 400 m. Thus, for low-frequency broadband signals, the long-range received signal will have a modal dispersion structure, and the dispersion structure will exhibit a modal dispersion interruption phenomenon. This phenomenon occurs because the modal eigenfunction has multiple nodes. For a deep source, when the source depth is at a node depth, the mode cannot be excited, forming a dispersion interruption. Therefore, the modal dispersion interruption corresponds to the first node position of the modal eigenfunction from bottom to top. The node effect leads to the modal dispersion interruption phenomenon, and the interruption band is determined by both the source depth and receiver depth.

When the receiver depth is close to the source depth, because the corresponding node frequencies are close, a dispersion interruption phenomenon appears in the modal time-frequency structure, with a large interruption frequency range. The upper limit of the interruption is determined by the smaller of the two depths, and the lower limit is determined by the larger of the two depths. Therefore, if a large interruption frequency is observed in the measured signal, and the interruption bands of each mode are relatively regular, if the receiver depth is known and the node frequency is further estimated, the other limit of the frequency band corresponds to the node frequency of the source depth.

When the receiver depth is significantly greater than the source depth, because the corresponding node frequencies are quite different, a frequency interruption phenomenon also appears in the modal time-frequency structure, but the interruption frequency range is small, and this interruption frequency is determined only by the source depth. Therefore, the source depth can be estimated based on this interruption frequency. At the same time, in this case, the high-frequency part of higher-order modes cannot be excited due to the limitation of the receiver depth, while the high-frequency part of lower-order modes can still be excited because of the larger vertical coverage of their eigenfunctions; therefore, only modes between higher and lower orders have dispersion interruption.

When the receiver depth is significantly smaller than the source depth, with sufficient signal-to-noise ratio, the modal dispersion structure is divided into multiple segments with several frequency interruptions in between, which is more complicated. It includes the first node frequency corresponding to the source depth, and multiple node frequencies corresponding to the receiver depth. Therefore, with the known receiver depth, the node frequencies caused by the receiver depth can be excluded by simulating the node frequencies, and the source depth can be estimated using the remaining node frequency.

Based on the above characteristics, the source depth can be estimated according to the modal dispersion interruption frequency combined with the modal eigenfunction under different receiver depth conditions. The receiver depth is used to exclude the nodes caused by the receiver, and then the source depth is estimated from the remaining interruption frequency.

This method has several advantages. Advantage one: there is no strict requirement for the receiver depth; it is applicable at any depth in the deep Arctic surface layer. Advantage two: since the dispersion interruption is located in the middle of the modal frequency band, ice attenuation does not affect the applicability of this method, whereas the source depth estimation method based on the modal upper frequency limit is affected by ice attenuation. Advantage three: this method is not affected by the double-duct sound speed profile, because the double-duct profile mainly affects the high-frequency part of the mode, causing crossing phenomena, while this method uses the lower frequency band without crossing problems. The crossing problem makes the upper limit of the mode difficult to determine, limiting the method based on the modal upper limit. In addition, this method can utilize hydrophones at depths exceeding 400 m, providing better source depth estimation results, because large-depth reception has a clear node dispersion interruption that is only affected by the source depth, and large-depth hydrophones have better concealment. A single hydrophone can achieve this function, making the entire underwater system easy to deploy and recover.

Regarding the multi-value problem where one interruption frequency corresponds to multiple source depths, the modal amplitude characteristic is used: the first interruption from large to small frequency corresponds to the first node depth of the modal eigenfunction from bottom to top, which is determined by the characteristic that the first wave of the Arctic sound speed profile eigenfunction from bottom to top has the maximum amplitude.